\documentclass[11pt,a4paper,dvipdfmx]{article}
\usepackage{jcappub,natbib}
\usepackage{braket}
\title{General parity-odd CMB bispectrum estimation}
\author[a,b]{Maresuke Shiraishi,}
\author[a,b]{Michele Liguori}
\author[c]{and James R. Fergusson}

\affiliation[a]{Dipartimento di Fisica e Astronomia ``G. Galilei'', \\ 
Universit\`a degli Studi di Padova, via Marzolo 8, I-35131, Padova, Italy}
\affiliation[b]{INFN, Sezione di Padova, \\ 
via Marzolo 8, I-35131, Padova, Italy}
\affiliation[c]{Centre for Theoretical Cosmology, \\ 
Department of Applied Mathematics and Theoretical Physics, \\ 
University of Cambridge, Wilberforce Road, Cambridge CB3 0WA, United Kingdom}

\abstract{%
We develop a methodology for estimating parity-odd bispectra in the cosmic microwave background (CMB). This is achieved through the extension of the original separable modal methodology to parity-odd bispectrum domains ($\ell_1 + \ell_2 + \ell_3 = {\rm odd}$). Through numerical tests of the parity-odd modal decomposition with some theoretical bispectrum templates, we verify that the parity-odd modal methodology can successfully reproduce the CMB bispectrum, without numerical instabilities. We also present simulated non-Gaussian maps produced by modal-decomposed parity-odd bispectra, and show the consistency with the exact results. Our new methodology is applicable to all types of parity-odd temperature and polarization bispectra.}

\begin{document}

\maketitle
\flushbottom

%%%%%%%%%%%%%%%%%%%%%%%%%%%%%%%%%%%%%%%
\section{Introduction}

Bispectrum estimation of the cosmic microwave background (CMB) is one of the most powerful ways to explore the non-Gaussianity of primordial fluctuations. While standard single-field slow-roll inflation predicts a tiny amount non-Gaussianity (NG) of 
the primordial curvature perturbations ~\cite{Acquaviva:2002ud,Maldacena:2002vr}, this is no longer true for a large number of extensions of the simplest inflationary paradigm (see e.g., refs.~\cite{Bartolo:2004if, Komatsu:2010hc} and references therein). Measurements of primordial NG thus provide a stringent test of the standard single-field slow roll scenario, and allow to put stringent constraints   on alternative models. The most stringent constraints on primordial NG to date have been obtained through bispectrum measurements of \textit{Planck} temperature data~\cite{Ade:2013ydc}. Future analyses, including correlations with E-mode polarization (and thus additional CMB bispectra of the type $\Braket{TTE}$, $\Braket{TEE}$ and $\Braket{EEE}$), will bring in further improvement on the current observational bounds \cite{Babich:2004yc, Yadav:2007rk}. 

All CMB NG searches so far have been focused on parity-even bispectra, in which the condition $\ell_1 + \ell_2 + \ell_3 = {\rm even}$ is enforced. This is because, as long as we consider the bispectrum of primordial curvature perturbations, parity cannot be broken, due to the spin-0 nature of the scalar mode. On the other hand, several interesting models predict bispectra generated by vector or tensor perturbations. In these cases the parity-even condition might have to be removed, since the vector or tensor modes can create parity-odd NG due to their spin dependence. For example, Early Universe models with some parity-violating or parity-odd sources, such as the gravitational and electromagnetic Chern-Simons actions \cite{Maldacena:2011nz, Soda:2011am, Barnaby:2012xt, Zhu:2013fja, Cook:2013xea}, or large-scale helical magnetic fields \cite{Caprini:2003vc, Kahniashvili:2005xe}, generate NG with sizable CMB bispectrum signals in parity-odd configurations ($\ell_1 + \ell_2 + \ell_3 = {\rm odd}$) \cite{Kamionkowski:2010rb, Shiraishi:2011st, Shiraishi:2012sn, Shiraishi:2013kxa}. Vector or tensor modes also induce B-mode polarization. B-mode bispectra can thus be useful to prove tensor NG \cite{Shiraishi:2013vha, Shiraishi:2013kxa}. At the same time, the parity-odd property of the B-mode field can generate $\ell_1 + \ell_2 + \ell_3 = {\rm odd}$ configurations in $\Braket{TTB}$, $\Braket{TEB}$, $\Braket{EEB}$ and $\Braket{BBB}$ bispectra, even when primordial NG has even parity. B-mode bispectra are also generated via secondary CMB lensing effects \cite{Lewis:2011fk}. These theoretical predictions motivate us to investigate the CMB signals in $\ell_1 + \ell_2 + \ell_3 = {\rm odd}$ triangles using observational data; hence, in this paper, we want to develop a general framework for parity-odd bispectrum estimation. 

CMB bispectrum estimation is generally aimed at measuring the so called non-linear parameter $f_{\rm NL}$. This can be done optimally 
by mean of the following estimator \cite{Komatsu:2008hk}:
\begin{eqnarray}
{\cal E} = \frac{1}{N^2}
\left[ \prod_{n=1}^3 \sum_{\ell_n m_n} \right]
\left(
\begin{array}{ccc}
\ell_1 & \ell_2 & \ell_3 \\
m_1 & m_2 & m_3
\end{array}
\right)
B_{\ell_1 \ell_2 \ell_3}    
\left[ \left( \prod_{n=1}^3 \frac{a_{\ell_n m_n}^{\rm O}}{C_{\ell_n}} \right)
 - 6  \frac{C_{\ell_1 m_1, \ell_2 m_2}}{C_{\ell_1} C_{\ell_2}} \frac{a_{\ell_3 m_3}^{\rm O}}{C_{\ell_3}} \right] ~, \label{eq:estimator_def} 
\end{eqnarray}
where $B_{\ell_1 \ell_2 \ell_3}$ is a theoretical template of the CMB angle-averaged bispectrum, $a_{\ell m}^{\rm O}$ are the observed CMB multipoles, $C_\ell$ is the CMB power spectrum, and $C_{\ell_1 m_1, \ell_2 m_2} = \Braket{a_{\ell_1 m_1}^{\rm G} a_{\ell_2 m_2}^{\rm G}}$ is the covariance matrix, obtained from simulated Gaussian maps $a_{\ell m}^{\rm G}$. Finally, 
\begin{eqnarray}
N^2 \equiv \sum_{\ell_1 \ell_2 \ell_3}
\frac{B_{\ell_1 \ell_2 \ell_3}^2 }{C_{\ell_1} C_{\ell_2} C_{\ell_3}}~, \label{eq:normalization}
\end{eqnarray}
is a normalization factor. The estimated $f_{\rm NL}$ parameter basically measures the degree of correlation between the theoretical template under study and the three-point function extracted from the data. 
The input $B_{\ell_1 \ell_2 \ell_3}$ and $C_\ell$, as well as the Monte Carlo simulations used for covariance matrix calculations, include all realistic experimental features such as instrumental beam, mask and noise. Note that the form of the estimator written above is derived under the ``diagonal covariance approximation'' i.e. we are replacing the general $C^{-1}$ filtering of the multipoles, where $C$ is the (in realistic experimental conditions non-diagonal) $a_{\ell m}$ covariance matrix, with a much simpler ${1/C_\ell}$ filtering. This in principle 
 implies some loss of optimality. In the context of \textit{Planck} data analysis, it was however shown ~\cite{Ade:2013ydc} that it is possible in practice to retain optimality using the simplified estimator above, provided the CMB map is pre-filtered by mean of a recursive inpainting technique. For this reason, we will work in diagonal covariance approximation throughout the rest of this work (in any case all of our derivation readily applies to the full-covariance expressions, by simply operating a $ {a_{\rm \ell m}/C_\ell} \rightarrow (C^{-1} a)_{\ell m}$ replacement). 
One important and well-known practical issue with the estimator of eq.~(\ref{eq:estimator_def}) is that its brute force numerical computation leads to ${\cal O}(\ell_{\rm max}^5)$ operations. This requires huge CPU time and, for the large $\ell_{\rm max}$ achieved in current and forthcoming observations, it makes a direct approach of this kind totally unfeasible. A similar issue also appears when simulating NG maps with a given bispectrum, using the following formula originally introduced in ref.~\cite{Smith:2006ud}:
\begin{eqnarray}
a_{\ell_1 m_1}^{\rm NG} =
\frac{1}{6}
\left[ \prod_{n=2}^3 \sum_{\ell_n m_n} 
\frac{a_{\ell_n m_n}^{{\rm G}*}}{C_{\ell_n}} \right] 
\left(
  \begin{array}{ccc}
  \ell_1 & \ell_2 & \ell_3 \\
  m_1 & m_2 & m_3
  \end{array}
 \right)
B_{\ell_1 \ell_2 \ell_3}~. \label{eq:almNG_def}
\end{eqnarray}
Such numerical issues can be solved if the theoretical bispectrum is given by a separable form in terms of $\ell_1$, $\ell_2$ and $\ell_3$. Using a general technique originally introduced in ref.~\cite{Komatsu:2003iq}, and often dubbed the KSW method, the estimator can then be written in terms of a separate product of filtered maps in pixel space, thus massively reducing the computational cost to  ${\cal O}(\ell_{\rm max}^3)$ operations. For the parity-even case, many bispectra can be directly written in separable form. In particular, the so called local, equilateral and orthogonal bispectra, encompassing a vast number of NG scenarios, can be described in terms of separable templates. The KSW approach is directly applicable in this case.
On the other hand, the parity-odd bispectra here under study originate from complicated spin and angle dependences in the vector or tensor NG, or coming from lensing effects, and hence they are generally given by a complex non-separable form. A natural way to circumvent this issue is to adopt the separable modal methodology, originally developed by \cite{Fergusson:2009nv, Fergusson:2010dm, Fergusson:2011sa} for parity-even templates, extending it to parity-odd bispectrum domains.
In the modal approach, a general non-separable bispectrum shape is expanded in terms of a suitably constructed, complete basis of separable bispectrum templates in harmonic or Fourier space. Provided we use enough templates in the expansion (with convergence speed depending on the choice of basis and the shape of the bispectrum to expand) we can always reproduce the starting template with as high as needed degree of accuracy, and the new expanded shape will be separable by construction.

In order to extend the methodology to parity-odd bispectra, we will have to introduce a new weight function to account for spin dependence, and redefine a reduced bispectrum which is not restricted by $\ell_1 + \ell_2 + \ell_3 = {\rm even}$.  After getting analytical expressions for our parity-odd estimator, we will numerically implement it for the three Early Universe models described in \cite{Shiraishi:2011st, Shiraishi:2012sn, Shiraishi:2013kxa}. This will allow us to confirm that the modal decomposition can be successfully applied to parity-odd bispectra. We also use the modal technique to produce NG maps including the bispectra under study.

This paper is organized as follows. In the next section, we summarize the original modal decomposition for the parity-even case. In section~\ref{sec:modal_odd}, we extend it to parity-odd models. In section~\ref{sec:example}, we discuss the numerical implementation of the method, showing several applications, and we draw our conclusions in the final section.

%%%%%%%%%%%%%%%%%%%%%%%%%%%%%%%%%%%%%%%
\section{Parity-even modal decomposition}\label{sec:modal_even}

Before moving to the discussion on the parity-odd case, we here summarize the original modal methodology, which was applied to the estimation of parity-even bispectra in WMAP and \textit{Planck} data \cite{Fergusson:2009nv, Fergusson:2010dm, Fergusson:2011sa, Ade:2013ydc}. 

The parity-even angle-averaged bispectrum, $B_{\ell_1 \ell_2 \ell_3}^{(e)}$, respects the following selection rules:  
\begin{eqnarray}
\ell_1 + \ell_2 + \ell_3 = {\rm even} ~, \ \  
|\ell_1 - \ell_2| \leq \ell_3 \leq \ell_1 + \ell_2 ~. \label{eq:selection_even}
\end{eqnarray}
Due to rotational invariance, all the physical information for parity-even bispectra is encoded in the so-called reduced bispectrum, $b_{\ell_1 \ell_2 \ell_3}^{(e)}$, defined by
\begin{eqnarray} 
B_{\ell_1 \ell_2 \ell_3}^{(e)} \equiv h_{\ell_1 \ell_2 \ell_3} b_{\ell_1 \ell_2 \ell_3}^{(e)} ~. \label{eq:CMB_red_bis_even}
\end{eqnarray}
In this expression, the model-independent weight function $h_{\ell_1 \ell_2 \ell_3}$, defined in terms of Wigner 3j-symbols as
\begin{eqnarray}
h_{\ell_1 \ell_2 \ell_3} 
\equiv \sqrt{\frac{(2\ell_1 + 1)(2\ell_2 + 1)(2\ell_3 + 1)}{4\pi}}
\left(
  \begin{array}{ccc}
  \ell_1 & \ell_2 & \ell_3 \\
  0 & 0 & 0 
  \end{array}
 \right) ~,
\end{eqnarray}
enforces the parity-even condition and the triangle inequalities (\ref{eq:selection_even}). 
As we were mentioning in the previous section, the reduced bispectrum arising from many inflationary models cannot be directly written in separable form, although this is an essential requirement for
$f_{\rm NL}$ estimation. The modal methodology is based on expanding the reduced bispectrum as a sum over separable basis templates ``bispectrum modes" $Q_{ijk}(\ell_1, \ell_2, \ell_3)$:
\begin{eqnarray}
\frac{v_{\ell_1} v_{\ell_2} v_{\ell_3}}{\sqrt{C_{\ell_1} C_{\ell_2} C_{\ell_3}} } b_{\ell_1 \ell_2 \ell_3}^{(e)} 
&=& 
\sum_{ijk} \alpha_{ijk}^Q Q_{ijk}(\ell_1, \ell_2, \ell_3) ~, \\ 
%----
Q_{ijk}(\ell_1, \ell_2, \ell_3) 
&\equiv& q_{\{i}(\ell_1) q_{j}(\ell_2) q_{k\}}(\ell_3)\label{eq:aQexp} \nonumber \\ 
&=& 
\frac{1}{6} 
 q_i(\ell_1) q_j(\ell_2) q_k(\ell_3) + 5~{\rm perms~in~}i,j,k ~,
\end{eqnarray}
where $C_\ell$ denotes the CMB angular power spectrum, and a weight function $v_\ell = (2 \ell + 1)^{1/6}$ is introduced to remove an overall $\ell^{-{1/2}}$ scaling in the bispectrum estimator functions; we also have used the generic notation $\{a,b,c\}$ to indicate permutations over the indices $a$,$b$,$c$: $A_{\{a} A_{b} A_{c\}} = \frac{1}{6}A_a A_b A_c + 5~{\rm perms~in~}a,b,c $.  The $q_p(\ell)$ quantities  
are built, starting from generic functions such as e.g. monomial of different degree, or cosine and sine plane waves, through an orthonormalization procedure, so that:
\begin{equation}
\Braket{ q_p(\ell), q_r(\ell) } = \delta_{pr} \; ,
\end{equation}
where $\delta_{pr}$ is a Kronecker delta, and the scalar product $\Braket{ \cdot , \cdot}$ between two (parity-even) functions of $\ell_1, \ell_2, \ell_3$ is defined in analogy to the correlator between two bispectrum shapes.
\begin{eqnarray}
\Braket{F, F'}_{e} \equiv \sum_{\ell_1 \ell_2 \ell_3} 
\left( \frac{h_{\ell_1 \ell_2 \ell_3}}{v_{\ell_1} v_{\ell_2} v_{\ell_3}} \right)^2
F(\ell_1, \ell_2, \ell_3) F'(\ell_1, \ell_2, \ell_3)~ \; .
\end{eqnarray}
Note that the $Q_{ijk}(\ell_1, \ell_2, \ell_3)$ templates, constructed with this procedure, form a complete but not orthonormal basis. We can obtain a basis of orthonormal templates $R_n(\ell_1, \ell_2, \ell_3)$ 
through  a rotation in bispectrum space:
\begin{equation}
R_n(\ell_1, \ell_2, \ell_3) = \sum_{p} \lambda_{np} Q_p(\ell_1, \ell_2, \ell_3) \; .
\end{equation}
In the last expression, for convenience we have labeled the triples $ijk$, defining a given $Q$ or $R$ template, by mean of a single index $n$. The rotation matrix $\lambda$ is a lower triangular matrix, obtained as:
\begin{equation}
\gamma^{-1} = \lambda^\top \lambda \; ,
\end{equation}
 where $\gamma$ is the matrix of scalar products of the templates $Q_n$:
\begin{equation}
\gamma_{pr} \equiv \Braket{ Q_p, Q_r}_e \; .
\end{equation}
 Then, the modal coefficients $\alpha_{n}^Q$ of eq.~(\ref{eq:aQexp}) are computed as:
\begin{equation}
\alpha_n^Q = \sum_{p} (\lambda^\top)_{np} \alpha_p^R \; ,
\end{equation}
where $\alpha_n^R$ are the coefficients of the bispectrum expansion using the orthonormal basis $R_n$ as
\begin{equation}
\frac{v_{\ell_1} v_{\ell_2} v_{\ell_3}}{\sqrt{C_{\ell_1} C_{\ell_2} C_{\ell_3}} } b_{\ell_1 \ell_2 \ell_3}^{(e)} = \sum_n \alpha_n^R R_n(\ell_1,\ell_2,\ell_3) \; .
\end{equation}
Due to orthonormality we get:
\begin{eqnarray}
\alpha_n^R = 
\Braket{ \frac{v_{\ell_1} v_{\ell_2} v_{\ell_3} b_{\ell_1 \ell_2 \ell_3}^{(e)}}{\sqrt{C_{\ell_1} C_{\ell_2} C_{\ell_3}}}, R_n(\ell_1,\ell_2,\ell_3) }_{e} ~. 
\end{eqnarray}
Using the expressions above, and the spin-0 Gaunt integral formula, 
\begin{eqnarray}
\int d^2 \hat{\bf n} Y_{\ell_1 m_1}(\hat{\bf n}) Y_{\ell_2 m_2}(\hat{\bf n}) Y_{\ell_3 m_3}(\hat{\bf n}) = h_{\ell_1 \ell_2 \ell_3} 
\left(
\begin{array}{ccc}
\ell_1 & \ell_2 & \ell_3 \\
m_1 & m_2 & m_3
\end{array}
\right)
~,
\end{eqnarray}
it was originally shown in ref.~\cite{Fergusson:2009nv} that the estimator (\ref{eq:estimator_def}) can be written as:
\begin{eqnarray}
{\cal E} = \frac{1}{N^2}
\sum_n \alpha_{n}^Q  \beta_{n}^Q \; . \label{eq:estimator_even} 
\end{eqnarray}
In the past expression the normalization can be conveniently written in terms of the $\alpha^R$ expansion coefficients as $N^2 = \sum_{n} (\alpha_{n}^R)^2$ (it can be easily verified using orthonormality of $R_n$). Moreover we have defined 
a quantity $\beta_n^Q$ related to the observational data as:
\begin{eqnarray}\label{eq:estimatoreven}
\beta_{n \leftrightarrow ijk}^{Q} \equiv \int d^2 \hat{\bf n} 
 \left[ M_{\{i}^{\rm O}(\hat{\bf n}) M_{j}^{\rm O}(\hat{\bf n}) M_{k\}}^{\rm O}(\hat{\bf n}) - 6 \Braket{M_{\{i}^{\rm G}(\hat{\bf n}) M_{j}^{\rm G}(\hat{\bf n})} M_{k\}}^{\rm O}(\hat{\bf n}) \right] \; ,
\end{eqnarray}
with 
\begin{eqnarray}
M_i(\hat{\bf n}) \equiv \sum_{\ell m} q_i(\ell) \frac{a_{\ell m} }{v_{\ell} \sqrt{C_\ell}}  Y_{\ell m}(\hat{\bf n}) ~,
\end{eqnarray}
being a map generated from given $a_{\ell m}$ and from the modal function $q_i(\ell)$. Likewise, the formula for generating NG maps, (\ref{eq:almNG_def}), can also be simplified as 
\begin{eqnarray}
a_{\ell m}^{\rm NG} 
=
\frac{\sqrt{C_{\ell}}}{6 v_{\ell}}
\sum_{n \leftrightarrow ijk} \alpha_{n }^Q 
\left[ q_{\{i}(\ell) 
\int d^2 \hat{\bf n} Y_{\ell m}(\hat{\bf n}) M_{j}^{\rm G}(\hat{\bf n}) M_{k\}}^{\rm G}(\hat{\bf n}) \right]^* ~.
\end{eqnarray}

Note how the use of the complete but not orthogonal basis $Q_n$ allows to develop a fast estimator by employing separability and reducing the total scaling from $\sim \ell_{\rm max}^5$ in the brute force approach, to $\sim \ell_{\rm max}^3$ operations in the 
 angular integral (\ref{eq:estimatoreven}). On the other hand, the use of the orthonormal basis $R_n$ provides a straightforward way to compute the estimator's normalization, as a well as a more transparent way to present several results, since the measured coefficients of the $R_n$ expansion are uncorrelated.

%%%%%%%%%%%%%%%%%%%%%%%%%%%%%%%%%%%%%%%%%%%%%%%%%%%%%%%%%%%%
\section{Parity-odd modal decomposition}\label{sec:modal_odd}

In this section we will discuss the extension of the modal methodology to parity-odd bispectra, which constitutes the original contribution of this work. A general parity-odd angle-averaged bispectrum, $B_{\ell_1 \ell_2 \ell_3}^{(o)}$, satisfies 
\begin{eqnarray}
\ell_1 + \ell_2 + \ell_3 = {\rm odd}~, \ \ |\ell_1 - \ell_2| \leq \ell_3 \leq \ell_1 + \ell_2 ~. \label{eq:selection_odd}
\end{eqnarray}
In the standard modal methodology, outlined in the previous section, these signals cannot be picked up because the averaged bispectrum (\ref{eq:CMB_red_bis_even}) is automatically restricted only to $\ell_1 + \ell_2 + \ell_3 = {\rm even}$ configurations, due to the parity-even selection rule of the weight function $h_{\ell_1 \ell_2 \ell_3}$. In order to build a parity-odd modal pipeline we first need to redefine the reduced bispectrum, and to introduce an $h_{\ell_1 \ell_2 \ell_3}$ function which includes spin-dependence and covers $\ell_1 + \ell_2 + \ell_3 = {\rm odd}$ configurations. This will naturally lead to a redefinition of the scalar product in parity-odd domains.

%#########################################
\subsection{Separable bispectrum}

Let us introduce a new $h_{\ell_1 \ell_2 \ell_3}$ function by including three spin values $x$, $y$ and $z$, satisfying $x + y + z = 0$:
\begin{eqnarray}
h_{\ell_1 \ell_2 \ell_3}^{\{xyz\}} 
&\equiv& \frac{1}{6} h_{\ell_1 \ell_2 \ell_3}^{x~y~z}  
+ {5~\rm perms~in~}x, y, z ~, \\ 
%----
h_{\ell_1 \ell_2 \ell_3}^{x~y~z}  
&\equiv& \sqrt{\frac{(2\ell_1 + 1)(2\ell_2 + 1)(2\ell_3 + 1)}{4\pi}}
\left(
  \begin{array}{ccc}
  \ell_1 & \ell_2 & \ell_3 \\
  x & y & z 
  \end{array}
 \right) ~.
\end{eqnarray}
Note that $h_{\ell_1 \ell_2 \ell_3}^{\{xyz\}}$ coincides with the parity-even geometrical function $h_{\ell_1 \ell_2 \ell_3}$ for $x = y = z = 0$, but it does not vanish in $\ell_1 + \ell_2 + \ell_3 = {\rm odd}$ for other specific $x$, $y$ and $z$. Using this function, we shall define a parity-odd reduced bispectrum, $b_{\ell_1 \ell_2 \ell_3}^{(o)}$, as
\begin{eqnarray}
B_{\ell_1 \ell_2 \ell_3}^{(o)} 
\equiv h_{\ell_1 \ell_2 \ell_3}^{\{xyz\}} b_{\ell_1 \ell_2 \ell_3}^{(o)}  ~,
\end{eqnarray}
where $b_{\ell_1 \ell_2 \ell_3}^{(o)}$ is symmetric under the permutation of $\ell_1$, $\ell_2$ and $\ell_3$ due to $h_{\ell_2 \ell_1 \ell_3}^{\{xyz\}} = (-1)^{\ell_1 + \ell_2 + \ell_3}h_{\ell_1 \ell_2 \ell_3}^{\{xyz\}}$ and $B_{\ell_2 \ell_1 \ell_3}^{(o)} = (-1)^{\ell_1 + \ell_2 + \ell_3}B_{\ell_1 \ell_2 \ell_3}^{(o)}$, and takes only purely imaginary values. A general bispectrum estimator takes into account all triangles for which $\ell_1, \ell_2, \ell_3 \geq 2$. In the parity-odd case we also have to satisfy the selection rules, eq.~(\ref{eq:selection_odd}). These conditions specify the allowed range of $x$, $y$ and $z$ as $(x,y,z) = (\pm 1, \pm 1, \mp 2)$ and its permutations. 

The modal decomposition of the parity-odd bispectrum reads
\begin{eqnarray}
\frac{v_{\ell_1} v_{\ell_2} v_{\ell_3}}{i \sqrt{C_{\ell_1} C_{\ell_2} C_{\ell_3}} } b_{\ell_1 \ell_2 \ell_3}^{(o)} 
=
\sum_{n \leftrightarrow ijk} \alpha_n^Q Q_n(\ell_1, \ell_2, \ell_3) ~, 
\end{eqnarray}
where the left-hand side becomes real thanks to a factor $1/i$. As $Q_n$, we can safely adopt the modal basis used in the original parity-even modal methodology, since $b_{\ell_1 \ell_2 \ell_3}^{(o)}$ is still a smooth function in the $\ell$-space tetrahedron as shown in figure~\ref{fig:Bis3D}. Hence, $\alpha_n^Q, Q_n \in \mathbb{R}$ holds also in this parity-odd case. The modal coefficient is given by
\begin{eqnarray}
\alpha_n^Q &=& \sum_p \gamma_{np}^{-1} 
\Braket{ \frac{v_{\ell_1} v_{\ell_2} v_{\ell_3} b_{\ell_1 \ell_2 \ell_3}^{(o)}}{i \sqrt{C_{\ell_1} C_{\ell_2} C_{\ell_3}}}, Q_p(\ell_1,\ell_2,\ell_3) }_o ~, \label{eq:alphaQ_odd}
\end{eqnarray}
where $\Braket{\cdot,\cdot}_o$ is the inner product in the parity-odd space, which will be defined in the next subsection, and $\gamma_{np} \equiv \Braket{Q_n, Q_p}_o$. In principle, this $\gamma_{np}$ does not coincide with the parity-even counterpart, even if we start from the same $Q_n$ templates, due to the difference of the inner product. 

Likewise, the decomposition with the orthonormal basis, $R_n = \sum_{p=0}^n \lambda_{np} Q_p$, can be expressed as 
\begin{eqnarray}
\frac{v_{\ell_1} v_{\ell_2} v_{\ell_3}}{i \sqrt{C_{\ell_1} C_{\ell_2} C_{\ell_3}} } b_{\ell_1 \ell_2 \ell_3}^{(o)}  
=
\sum_{n \leftrightarrow ijk} \alpha_n^R R_n(\ell_1, \ell_2, \ell_3) ~,
\end{eqnarray}
where the rotation matrix $\lambda_{np}$ given by $(\gamma^{-1})_{np} = \sum_r (\lambda^\top)_{nr} \lambda_{rp}$ again differs from the parity-even counterpart. Therefore the $R_n$ orthonormal basis will also be different.

%#####################################################
\subsection{Inner product}

The scaling behavior of $h_{\ell_1 \ell_2 \ell_3}^{\{xyz\}}$ is similar to $h_{\ell_1 \ell_2 \ell_3}$, hence we can define the inner product between two real functions, defined on a parity-odd domain, in total analogy to the parity-even case:  
\begin{eqnarray}\label{eq:scalodd}
\Braket{F, F'}_o \equiv \sum_{\ell_1 + \ell_2 + \ell_3 = {\rm odd}} 
\left( \frac{h_{\ell_1 \ell_2 \ell_3}^{\{xyz\}}}{v_{\ell_1} v_{\ell_2} v_{\ell_3}} \right)^2
F(\ell_1, \ell_2, \ell_3) F'(\ell_1, \ell_2, \ell_3)~.
\end{eqnarray}
Here note that we must restrict the summation range as $\ell_1 + \ell_2 + \ell_3 = {\rm odd}$ by hand since $h_{\ell_1 \ell_2 \ell_3}^{\{xyz\}}$ does not vanish also in $\ell_1 + \ell_2 + \ell_3 = {\rm even}$.\footnote{Note that the same scalar product includes both parity-odd and parity-even configurations, so in principle it can be used also to expand parity-even bispectra.} 

In order to obtain the $\gamma$ and $\lambda$ matrices, which are necessary to generate the modal expansion, we have to compute a large number of scalar products between basis templates (more precisely, due to symmetry of $\gamma$, we need to compute $n_{\rm max}(n_{\rm max}+1)/2$ products, where $n_{\rm max}$ is the number of templates used in the numerical expansions, which is typically of order $10^2-10^3$, depending on the bispectrum we want to expand) . This poses a practical problem, 
since the number of operations in a brute-force approach scales like $\sim \ell_{\max}^3$ for each element of $\gamma$.

Fortunately, when the functions $F(\ell_1, \ell_2, \ell_3)$ and $F'(\ell_1, \ell_2, \ell_3)$ can be written in separable forms as $F(\ell_1, \ell_2, \ell_3) = f_{\{i}(\ell_1) f_{j}(\ell_2) f_{k\}}(\ell_3)$ and $F'(\ell_1, \ell_2, \ell_3) = f'_{\{i'}(\ell_1) f'_{j'}(\ell_2) f'_{k'\}}(\ell_3)$  (which is by construction always the case for the basis templates), we can follow an approach similar to the one introduced in ref.~\cite{Smith:2006ud}, in order to speed up the numerical computation of eq.~(\ref{eq:scalodd}). We start by rewriting the scalar product in terms of angular integrals: 
\begin{eqnarray}
\Braket{F, F'}_o &=& 8\pi^2
\int_{-1}^1 d \mu  \left[
{}_{\{-x}{\cal Y}_{\{i\{i' \{x}^{(o)}~ {}_{-y}{\cal Y}_{j j' y}^{(e)} ~ {}_{-z\}}{\cal Y}_{k\} k'\} z\}}^{(e)} 
%\nonumber \\ 
%---
%&& 
+  
{}_{\{-x}{\cal Y}_{\{i\{i' \{x }^{(e)}~ {}_{-y}{\cal Y}_{j j' y}^{(o)} ~ {}_{-z\}}{\cal Y}_{k\} k'\} z\}}^{(e)} \right. \nonumber \\ 
%---
&&\left.\qquad  + 
{}_{\{-x}{\cal Y}_{\{i\{i' \{x }^{(e)}~ {}_{-y}{\cal Y}_{j j' y}^{(e)} ~ {}_{-z\}}{\cal Y}_{k\} k'\} z\}}^{(o)} 
%\nonumber \\ 
%---
%&& 
+ 
{}_{\{-x}{\cal Y}_{\{i\{i' \{x}^{(o)}~ {}_{-y}{\cal Y}_{j j' y}^{(o)} ~ {}_{-z\}}{\cal Y}_{k\} k'\} z\}}^{(o)} \right] ~, \label{eq:prod_odd_sep}
\end{eqnarray}
where we have introduced a map depending on spin and parity as 
\begin{eqnarray}
{}_x{\cal Y}_{ii' y}^{(o/e)}(\mu) &\equiv& \sum_{\ell = {\rm odd/ even}}\sqrt{\frac{2\ell +1}{4\pi}} \frac{f_i(\ell) f'_{i'}(\ell)}{v_\ell^2} {}_{x}\lambda_{\ell y}(\mu) ~,
\end{eqnarray}
with ${}_{x}Y_{\ell y}(\hat{\bf n}) \equiv {}_x \lambda_{\ell y}(\mu) e^{iy\phi}$. To obtain this form, we have used the spin-dependent Gaunt integral,  
\begin{eqnarray}
\int d^2 \hat{\bf n} {}_{-s_1}Y_{\ell_1 m_1}(\hat{\bf n}) {}_{-s_2} Y_{\ell_2 m_2}(\hat{\bf n}) {}_{-s_3}Y_{\ell_3 m_3}(\hat{\bf n}) 
= h_{\ell_1 \ell_2 \ell_3}^{s_1 s_2 s_3} 
\left(
  \begin{array}{ccc}
  \ell_1 & \ell_2 & \ell_3 \\
  m_1 & m_2 & m_3 
  \end{array}
 \right)
  ~, \label{eq:gaunt_spin}
\end{eqnarray}
and the fact that the $\phi$ dependence in each integrand cancels out due to the spin conservation, namely, $x + y + z = 0$. Since ${}_{x}\lambda_{\ell y}(\mu)$ behaves like the Legendre polynomial, the $\mu$ integrals can be numerically estimated with a high degree of accuracy through Gauss-Legendre quadrature. The total number of angular integrals to compute is $6^4 \times 4$ at most and hence eq.~(\ref{eq:prod_odd_sep}) reduces the numerical operations from ${\cal O}(\ell_{\rm max}^3)$ to ${\cal O}(10^3 \times  \ell_{\rm max})$, once the set of all ${}_x{\cal Y}_{ii' y}^{(o/e)}$'s is pre-computed. As mentioned at beginning, the reduced form (\ref{eq:prod_odd_sep}) can then be used to speed up the calculation of $\gamma_{np}$, which is explicitly written down as 
\begin{eqnarray} 
\gamma_{np} &=& \Braket{Q_{n \leftrightarrow ijk}, Q_{p \leftrightarrow i'j'k'}}_o \nonumber \\ 
&=& 8\pi^2
\int_{-1}^1 d \mu 
\left[ 
{}_{\{-x}\zeta_{\{i\{i' \{x}^{(o)}~ {}_{-y}\zeta_{j j' y}^{(e)} ~ {}_{-z\}}\zeta_{k\} k'\} z\}}^{(e)} 
+ {}_{\{-x}\zeta_{\{i\{i' \{x}^{(e)}~ {}_{-y}\zeta_{j j' y}^{(o)} ~ {}_{-z\}}\zeta_{k\} k'\} z\}}^{(e)} \right. \nonumber \\ 
&&\left.\qquad  
+ {}_{\{-x}\zeta_{\{i\{i' \{x}^{(e)}~ {}_{-y}\zeta_{j j' y}^{(e)} ~ {}_{-z\}}\zeta_{k\} k'\} z\}}^{(o)} 
+ {}_{\{-x}\zeta_{\{i\{i' \{x}^{(o)}~ {}_{-y}\zeta_{j j' y}^{(o)} ~ {}_{-z\}}\zeta_{k\} k'\} z\}}^{(o)}  
\right]  ~, 
\end{eqnarray}
with 
\begin{eqnarray}
{}_x\zeta_{ii' y}^{(o/e)}(\mu) &\equiv& \sum_{\ell = {\rm odd} / {\rm even}}\sqrt{\frac{2\ell +1}{4\pi}} 
\frac{q_i(\ell) q_{i'}(\ell)}{v_\ell^2} {}_{x}\lambda_{\ell y}(\mu) ~.
\end{eqnarray}

%#################################################### 
\subsection{Fast estimator and non-Gaussian map}

Using the new parity-odd modal decomposition, and the spin-dependent Gaunt integral (\ref{eq:gaunt_spin}), the estimator (\ref{eq:estimator_def}) can be written as:
\begin{eqnarray}
{\cal E} = 
\frac{i}{N^2}
\sum_n \alpha_{n}^Q 
\beta_{n}^Q
~, \label{eq:estimator_odd}
\end{eqnarray}
where the normalization becomes negative because $b_{\ell_1 \ell_2 \ell_3}^{(o)}$ takes only purely imaginary values, so that $N^2 = - \sum_{np} \alpha_{n}^Q \gamma_{np} \alpha_{p}^Q$, and 
\begin{eqnarray}
\beta_{n \leftrightarrow ijk}^{Q} 
&\equiv& 
\int d^2 \hat{\bf n} 
 \left[ {}_{\{-x}M_{\{i}^{{\rm O}(o)} ~  {}_{-y}M_{j}^{{\rm O}(e)} ~ {}_{-z\}}M_{k\}}^{{\rm O}(e)} 
- 6 \Braket{{}_{\{-x}M_{\{i}^{{\rm G}(o)}~ {}_{-y}M_{j}^{{\rm G}(e)}} {}_{-z\}}M_{k\}}^{{\rm O}(e)} 
\right. \nonumber \\ 
&& \left.\quad+ 
{}_{\{-x}M_{\{i}^{{\rm O}(e)} ~  {}_{-y}M_{j}^{{\rm O}(o)} ~ {}_{-z\}}M_{k\}}^{{\rm O}(e)} 
- 6 \Braket{{}_{\{-x}M_{\{i}^{{\rm G}(e)}~ {}_{-y}M_{j}^{{\rm G}(o)}} {}_{-z\}}M_{k\}}^{{\rm O}(e)} 
\right. \nonumber \\ 
&&\left.\quad + 
{}_{\{-x}M_{\{i}^{{\rm O}(e)} ~  {}_{-y}M_{j}^{{\rm O}(e)} ~ {}_{-z\}}M_{k\}}^{{\rm O}(o)} 
- 6 \Braket{{}_{\{-x}M_{\{i}^{{\rm G}(e)}~ {}_{-y}M_{j}^{{\rm G}(e)}} {}_{-z\}}M_{k\}}^{{\rm O}(o)} 
\right. \nonumber \\ 
&&\left.\quad+ 
{}_{\{-x}M_{\{i}^{{\rm O}(o)} ~  {}_{-y}M_{j}^{{\rm O}(o)} ~ {}_{-z\}}M_{k\}}^{{\rm O}(o)} 
- 6 \Braket{{}_{\{-x}M_{\{i}^{{\rm G}(o)}~ {}_{-y}M_{j}^{{\rm G}(o)}} {}_{-z\}}M_{k\}}^{{\rm O}(o)} 
\right] ~. 
\end{eqnarray}
In the parity-odd case, a map generated from $a_{\ell m}$ involves the spin and parity dependence:
\begin{eqnarray}
{}_x M_i^{(o/e)}(\hat{\bf n}) \equiv \sum_{\ell = {\rm odd / even}} \sum_m q_i(\ell) \frac{a_{\ell m} }{v_{\ell} \sqrt{C_\ell}}  {}_x Y_{\ell m}(\hat{\bf n}) ~.
\end{eqnarray} 
Equation~(\ref{eq:estimator_odd}) requires a comparable number of numerical operations with respect to the original parity-even modal methodology, namely ${\cal O}(\ell_{\rm max}^3)$. The estimator can be written as usual 
 in terms of the orthonormal basis as 
\begin{eqnarray}
{\cal E} = 
\frac{i}{N^2}
\sum_n \alpha_{n}^R
\beta_{n}^R
~,
\end{eqnarray}
with $N^2 = - \sum_{n} (\alpha_{n}^R)^2$.

In the same manner, the form of the simulated NG map (\ref{eq:almNG_def}) is also reduced to 
\begin{eqnarray}
a_{\ell m}^{\rm NG} 
&=& 
\frac{i \sqrt{C_{\ell}}}{6 v_{\ell}}
\sum_{n \leftrightarrow ijk} \alpha_{n}^Q 
 q_{\{i}(\ell) 
\int d^2 \hat{\bf n} 
{}_{\{-x}Y_{\ell m}^*
\nonumber \\ 
&&\times 
\begin{cases}
\left[{}_{-y}M_{j}^{{\rm G} (o)}~ {}_{-z\}} M_{k\}}^{{\rm G} (o)}
+ {}_{-y}M_{j}^{{\rm G} (e)}~ {}_{-z\}} M_{k\}}^{{\rm G} (e)}
\right]^* 
& (\ell = {\rm odd}) \\ 
\left[{}_{-y}M_{j}^{{\rm G} (o)}~ {}_{-z\}} M_{k\}}^{{\rm G} (e)}
+ {}_{-y}M_{j}^{{\rm G} (e)}~ {}_{-z\}} M_{k\}}^{{\rm G} (o)}
\right]^* 
& (\ell = {\rm even})
\end{cases} . 
\end{eqnarray}

%%%%%%%%%%%%%%%%%%%%%%%%%%%%%%%%%%%%%%% 
\section{Practical application} \label{sec:example}

In this section, we shall present numerical results from the modal decompositions of parity-odd temperature bispectra arising from some interesting Early Universe models. In this analysis, without loss of generality, we choose the spin set in the geometrical function as $(x,y,z) = (1,1,-2)$. 

%################################### 
\subsection{Theoretical models creating parity-odd bispectra}  

\begin{figure}
  \begin{center}
    \includegraphics[width =1\textwidth]{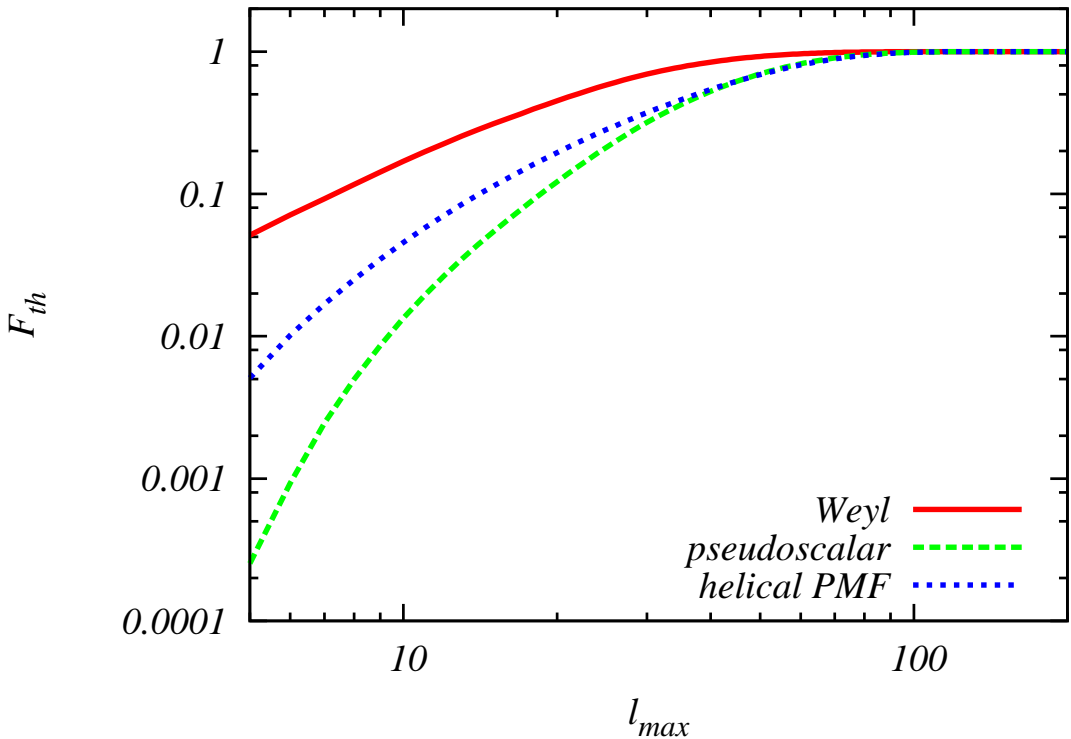}
  \end{center}
  \caption{Signal-to-noise ratios of the parity-odd temperature bispectrum in the Weyl (red solid line), pseudoscalar (green dashed line) and helical PMF (blue dotted line) models, respectively. The values are normalized at $\ell_{\rm max} = 200$.}
  \label{fig:SN2}
\end{figure}

Parity-odd signatures in the CMB bispectra \cite{Shiraishi:2011st, Shiraishi:2013kxa, Shiraishi:2012sn} can arise when we consider Early Universe models including Chern-Simons like actions in Weyl gravity ($f(\phi) \widetilde{W}W^2$) \cite{Maldacena:2011nz, Soda:2011am} or inflationary models with rolling pseudoscalars ($\phi \widetilde{F}F$) \cite{Barnaby:2012xt, Cook:2013xea}, or when we have primordial parity-odd objects, like helical primordial magnetic fields (PMFs) (see e.g., refs.~\cite{Caprini:2003vc, Kahniashvili:2005xe}). These signatures arise not from scalar, due to their spin-0 nature, but from tensor perturbations. Tensor-mode temperature signals are transferred to CMB temperature fluctuations via the Integrated Sachs-Wolfe (ISW) effect on large scales, while polarization signals appear on smaller scales due to Thomson scattering effects \cite{Pritchard:2004qp}. Accordingly, the temperature bispectra induced from the above NG sources can be large for $\ell \lesssim 100$. 

Figure~\ref{fig:SN2} describes signal-to-noise ratios of the parity-odd temperature bispectra estimated  for the three theoretical scenarios mentioned above (i.e., the so called Weyl, pseudoscalar and helical PMF models \cite{Shiraishi:2011st, Shiraishi:2013kxa, Shiraishi:2012sn}) by mean of a brute force approach. The Fisher matrices for these models can be written, using the scalar product (\ref{eq:scalodd}), as:
\begin{eqnarray} 
F_{\rm th} = \frac{1}{6} \Braket{ \frac{v_{\ell_1} v_{\ell_2} v_{\ell_3} b_{\ell_1 \ell_2 \ell_3}^{(o)}}{i \sqrt{C_{\ell_1} C_{\ell_2} C_{\ell_3}}}, 
\frac{v_{\ell_1} v_{\ell_2} v_{\ell_3} b_{\ell_1 \ell_2 \ell_3}^{(o)}}{i \sqrt{C_{\ell_1} C_{\ell_2} C_{\ell_3}}} }_o ~. \label{eq:SN2}
\end{eqnarray}
As shown in this figure, all the signal-to-noise ratios are saturated for $\ell_{\rm max} \gtrsim 100$, due to the end of the ISW enhancement of the tensor mode.

\begin{figure}[t]
  \begin{tabular}{cc}
    \begin{minipage}{0.5\hsize}
  \begin{center}
    \includegraphics[width = 1\textwidth]{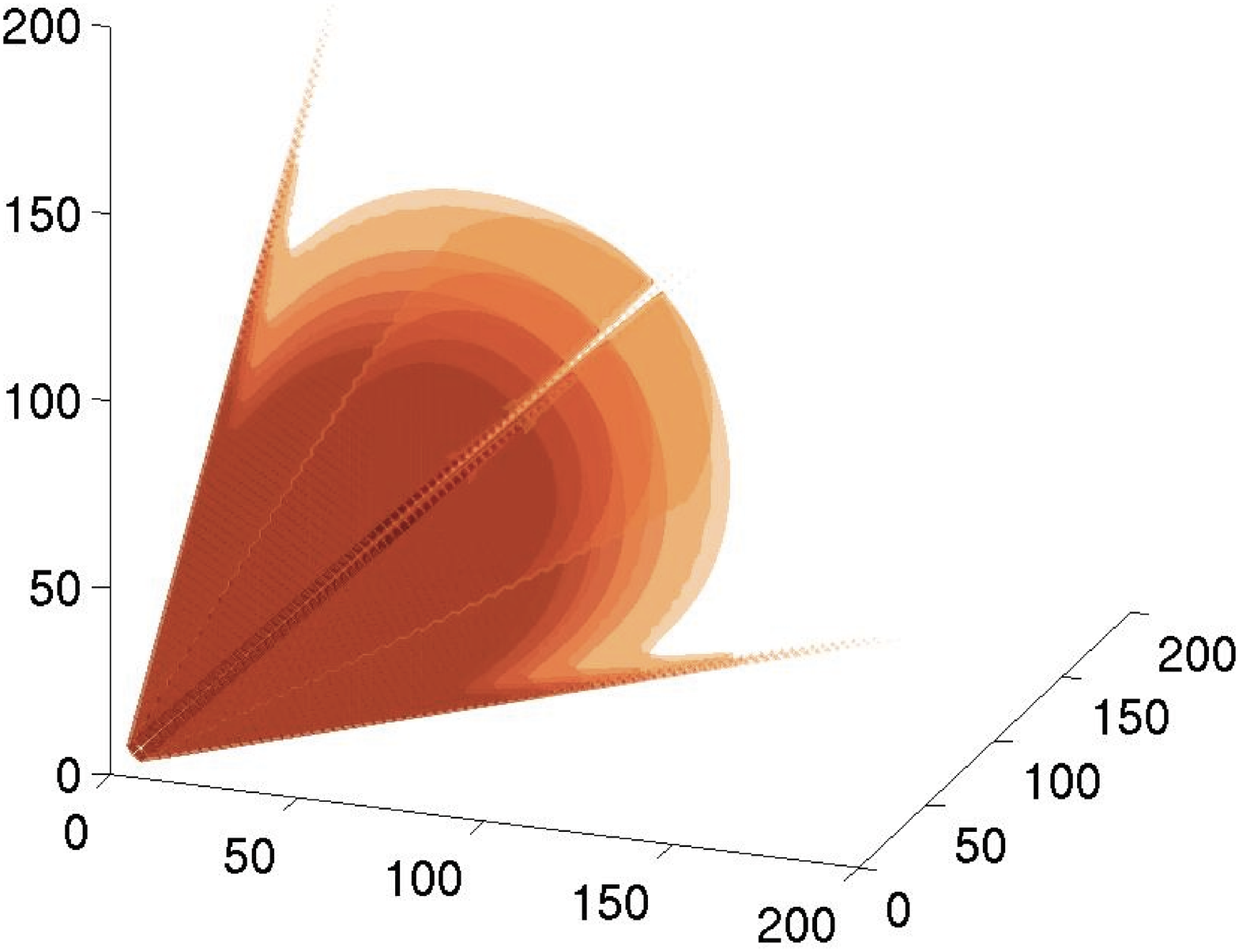}
  \end{center}
\end{minipage}
\begin{minipage}{0.5\hsize}
  \begin{center}
    \includegraphics[width = 1\textwidth]{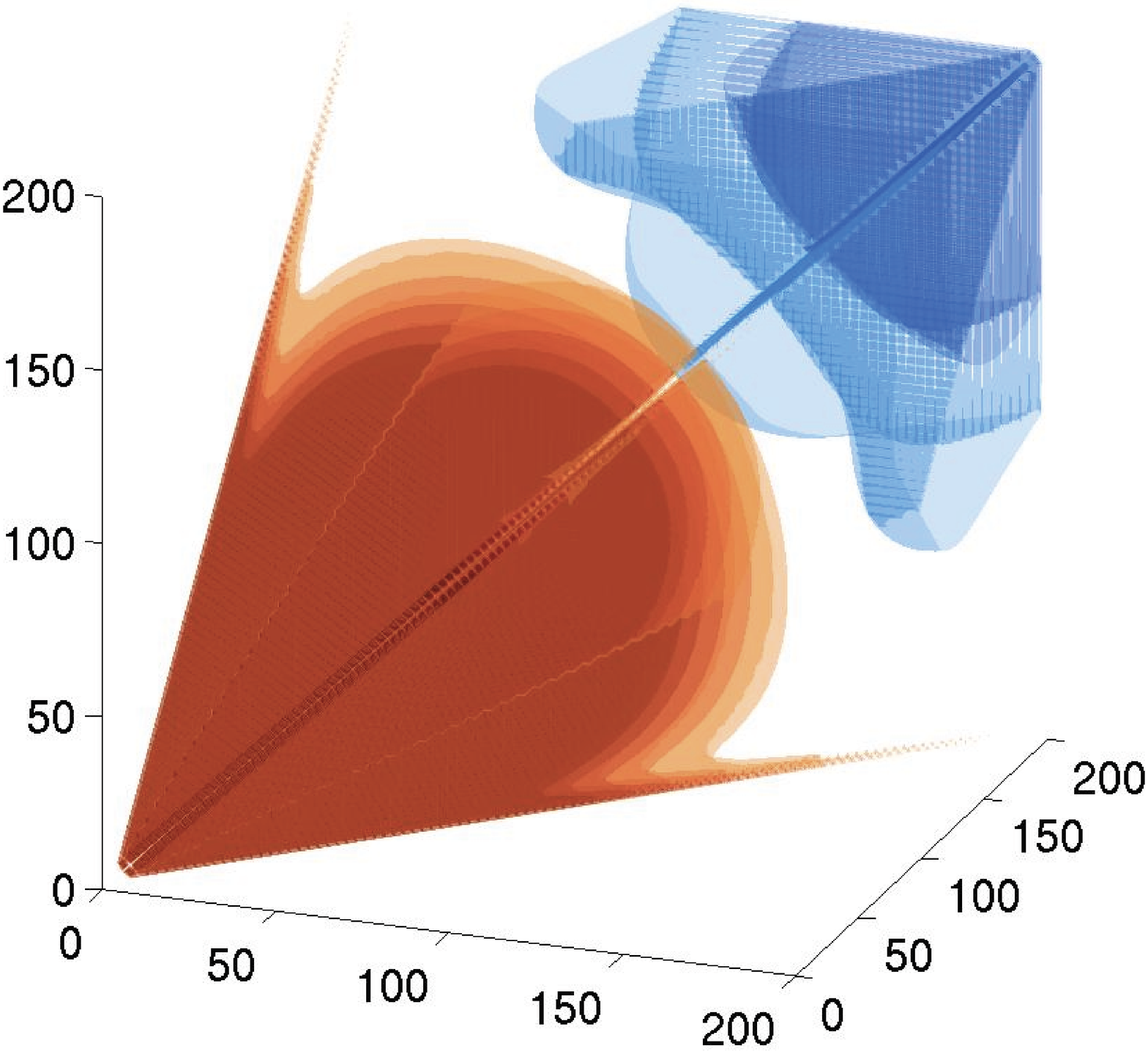}
  \end{center}
\end{minipage}
\end{tabular}
\\
  \begin{tabular}{c}
    \begin{minipage}{1.0\hsize}
  \begin{center}
    \includegraphics[width = 0.5\textwidth]{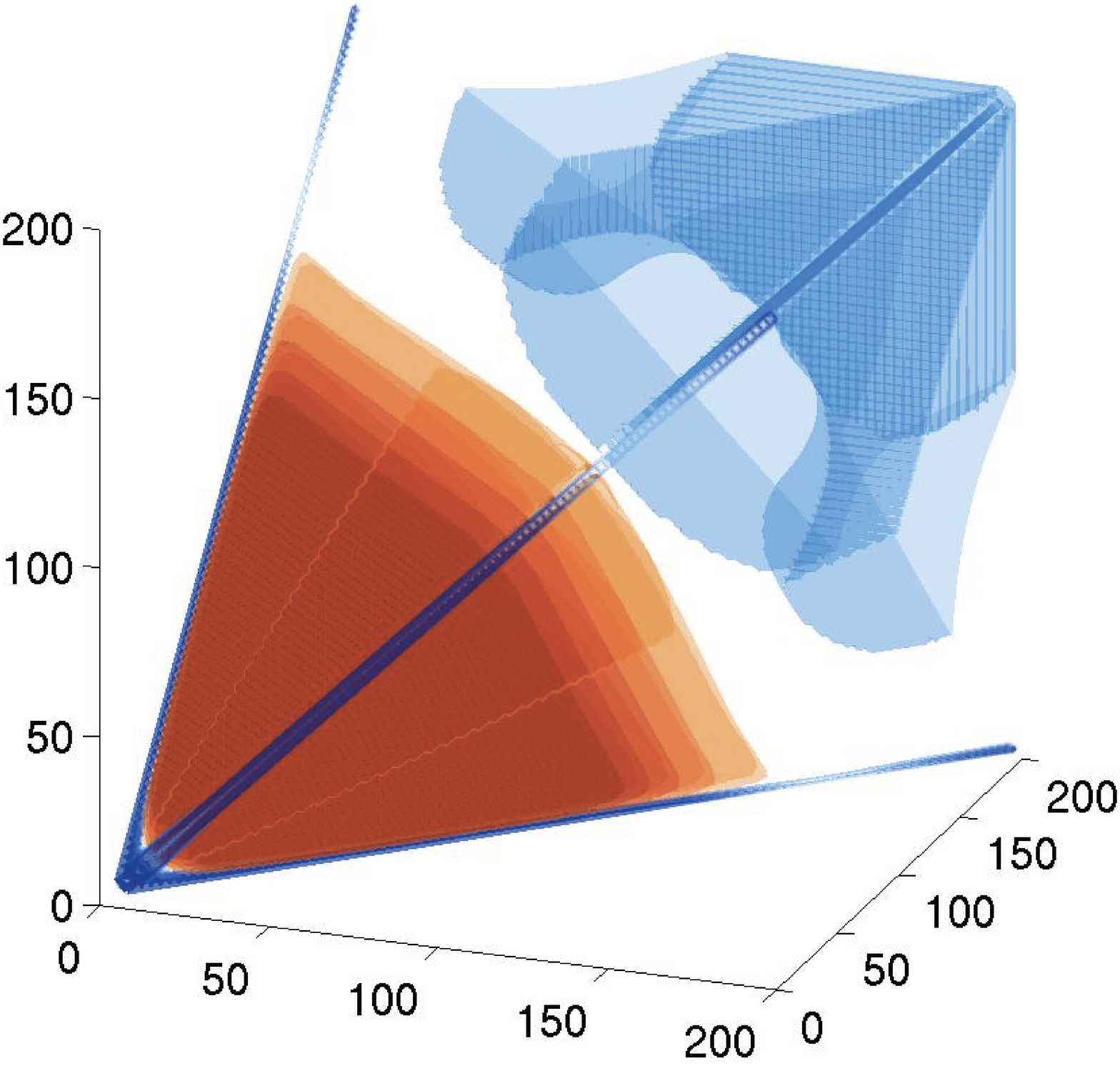}
  \end{center}
\end{minipage}
\end{tabular}
  \caption{Three-dimensional representation of Weyl (left top panel), pseudoscalar (right top panel) and helical PMF (bottom panel) reduced bispectra in the tetrahedral domain \eqref{eq:selection_odd}. In order to remove an $\ell^{-4}$ scaling, we rescale the shapes using a constant Sachs-Wolfe template \cite{Fergusson:2009nv}.}
\label{fig:Bis3D} 
\end{figure}

Three-dimensional representations of these bispectra are described in figure~\ref{fig:Bis3D}. We can clearly see from the figure that the Weyl and pseudoscalar bispectra peak in the equilateral limit, while the helical PMF bispectrum peaks on the squeezed configurations. Note that these parity-odd bispectra vanish if two or three multipoles take the same values, i.e., $B_{\ell \ell \ell} = B_{\ell \ell \ell'} = B_{\ell \ell' \ell} = 0$. 

In the following subsections, we will obtain modal decompositions for the three models above, and compare their performances to the brute-force approach. For safety and further accuracy, we will expand the bispectrum signals up to $\ell = 200$, although the signal-to-noise ratios almost converge for $\ell_{\rm max} \sim 100$. For convenience in the comparisons, we will fix the normalization of the bispectrum templates in such a way as to get $F_{\rm th}(\ell_{\rm max} = 200) = 1$, as described in figure~\ref{fig:SN2}.
\footnote{These bispectrum amplitudes in the Weyl, pseudoscalar and helical PMF model are obtained when $\Lambda r^{-2} = 1.1 \times 10^8$~GeV, $X = 9.7 \times 10^5$ and ${\cal B}_{1}^{1/3} B_{1}^{2/3} = 2.3$~nG, respectively, where $\Lambda r^{-2}$ is a combination of an energy scale of the dual cubic action and a tensor-to-scalar ratio in the Weyl gravity, $X$ is a coupling parameter of a pseudoscalar, and ${\cal B}_{1}^{1/3} B_{1}^{2/3}$ is a combination of the helical and non-helical magnetic field strengths per 1 Mpc generated at the GUT epoch \cite{Shiraishi:2011st, Shiraishi:2013kxa, Shiraishi:2012sn}. These parameter values will be detected at 68\% CL through the parity-odd bispectrum estimation with noiseless large-scale ($\ell \lesssim 200$) data given by WMAP or {\it Planck}.}

%###################################
\subsection{Modal decomposition}

For $(x,y,z) = (1,1,-2)$, $\gamma_{np}$ can explicitly written as
\begin{eqnarray}
\gamma_{np}
&=& \frac{8 \pi^2}{3}
\int_{-1}^1 d\mu
~{}_{-1}\zeta_{\{i \{i' 1}^{(o)}
\left( {}_{-1}\zeta_{j j' 1}^{(o)}% (\hat{n}) 
~{}_{2}\zeta_{k\} k'\} -2}^{(o)}%(\hat{n}) 
+ 2 {}_{-1}\zeta_{j j' -2}^{(o)}% (\hat{n}) 
~{}_{2}\zeta_{k\} k'\} 1}^{(o)}%(\hat{n}) 
\right)  \nonumber \\ 
%------
&&+ \frac{16 \pi^2}{3}
\int_{-1}^1 d\mu
~{}_{-1}\zeta_{\{i \{i' 1}^{(o)}
\left( {}_{-1}\zeta_{j j' 1}^{(e)}% (\hat{n}) 
~{}_{2}\zeta_{k\} k'\} -2}^{(e)}%(\hat{n}) 
+ {}_{-1}\zeta_{j j' -2}^{(e)}% (\hat{n}) 
~{}_{2}\zeta_{k\} k'\} 1}^{(e)}%(\hat{n}) 
\right)  \nonumber \\ 
%------
&& 
+ \frac{16 \pi^2}{3}
\int_{-1}^1 d\mu 
\left( 
{}_{-1}\zeta_{\{i \{i' -2}^{(o)}% (\hat{n}) 
~{}_{-1}\zeta_{j j' 1}^{(e)}% (\hat{n}) 
~{}_{2}\zeta_{k\} k'\} 1}^{(e)}%(\hat{n}) 
% \nonumber \\ 
%------
%&&
+ {}_{2}\zeta_{\{i \{i' 1}^{(o)}% (\hat{n}) 
~{}_{-1}\zeta_{j j' 1}^{(e)}% (\hat{n}) 
~{}_{-1}\zeta_{ k\} k'\} -2}^{(e)}%(\hat{n}) 
\right)
 \nonumber \\ 
%-----
&&+ 
\frac{8 \pi^2}{3}
\int_{-1}^1 d\mu  
~{}_{2}\zeta_{\{i \{i' -2}^{(o)}% (\hat{n}) 
~{}_{-1}\zeta_{j j' 1}^{(e)}% (\hat{n}) 
~{}_{-1}\zeta_{ k\} k'\} 1}^{(e)}%(\hat{n}) 
~. 
\end{eqnarray}
Starting from this $\gamma$, we can calculate the expansion coefficients $\alpha_n^Q$ by use of eq.~(\ref{eq:alphaQ_odd}) and this $\gamma_{np}$, and rotate to the orthonormal basis to get $\alpha_n^R$.  We implemented a stable numerical algorithm that perform these operations using about $200$ modes in $5$ CPU hours. 

\begin{figure}
  \begin{center}
    \includegraphics[width =1\textwidth]{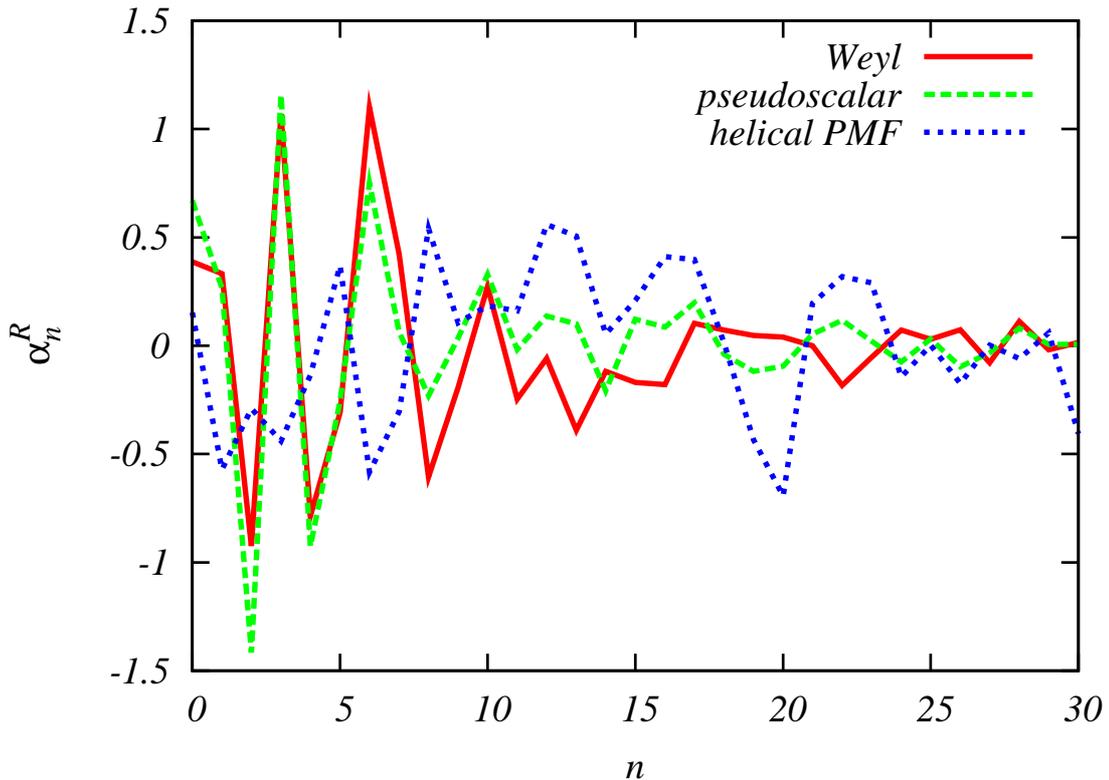}
  \end{center}
  \caption{Modal coefficients in the orthonormal basis, $\alpha_n^R$, in the Weyl (red solid line), pseudoscalar (green dashed line) and helical PMF (blue dotted line) models, respectively. Here we decompose with the polynomial eigenfunctions and an additional function ($n=1$) sensitive to the squeezed-limit signals.}
  \label{fig:alphaR}
\end{figure}

Figure~\ref{fig:alphaR} depicts the spectrum of mode coefficients as a function of mode number each $n$. We plot as usual the $\alpha^R$ coefficients, related to the orthonormal expansion. The interpretation of the results in the rotated space is in fact more straightforward due to the fact that, as a consequence of orthonormality of the basis templates, the expansion coefficients are uncorrelated. 
Here, we have decomposed using an hybrid basis involving  polynomial eigenfunctions and a special mode function enhanced at the squeezed limit (this is the same basis used for the analysis 
of \textit{Planck} data~\cite{Ade:2013ydc} using the parity-even pipeline; here we use however a smaller number of modes, since we work at lower $\ell_{\rm max}$, which allows for more rapid convergence). In the chosen mode ordering, the squeezed mode is located at $n=1$. In the helical PMF case, the $n=1$ mode is relatively large compared with the other modes due to the squeezed-limit amplification, while, as expected, it is not prominent in the equilateral-type bispectra in the Weyl and pseudoscalar cases. We find that $\alpha_n^R$'s in these equilateral bispectra converge more rapidly compared with the squeezed-type one. 

\begin{figure}
  \begin{center}
    \includegraphics[width =1\textwidth]{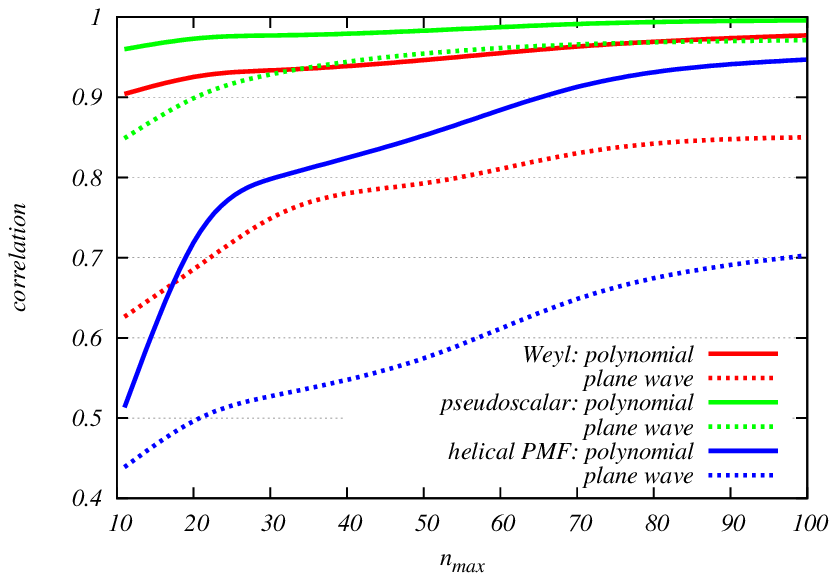}
  \end{center}
  \caption{Correlations between exact bispectra and modal-decomposed ones up to $n_{\rm max}$, in the Weyl (red lines), pseudoscalar (green lines) and helical PMF (blue lines) models, respectively. Here, we compare the differences in convergence between the polynomial + squeezed mode (solid lines) and the plane-wave + squeezed mode (dotted lines) decompositions.}
  \label{fig:corr_coeff}
\end{figure}

The convergence of the expansion is assessed via computation of the correlation between the full starting theoretical bispectrum and the modal-decomposed one. Using the orthonormal basis expansion, it is easy to verify that 
the correlation between the two is expressed as $F_{{\rm th} R} / \sqrt{ F_{\rm th} F_{R} }$, where each Fisher matrix is given by eq.~(\ref{eq:SN2}) and 
\begin{eqnarray}
F_{{\rm th} R} &=& \frac{1}{6} \Braket{\frac{v_{\ell_1} v_{\ell_2} v_{\ell_3} b_{\ell_1 \ell_2 \ell_3}^{(o)}}{i \sqrt{C_{\ell_1} C_{\ell_2} C_{\ell_3}}} , \sum_{n=0}^{n_{\rm max}} \alpha_n^R R_n(\ell_1, \ell_2, \ell_3)}_o ~, \\ 
%-------
F_{R} &=& \frac{1}{6} \sum_{n=0}^{n_{\rm max}} (\alpha_n^R)^2 ~.
\end{eqnarray}
Figure \ref{fig:corr_coeff} shows numerical results of the correlations  obtained from the modal decompositions as a function of $n_{\rm max}$, where $n_{\rm max}$ denotes the number of modes at which we truncate our numerical expansions. For this analysis, we have adopted both the hybrid polynomial-squeezed basis described above and an additional plane-wave decomposition, also equipped with the $n=1$ squeezed mode. Like the polynomial basis, also the plane-wave expansion was already adopted in the context of WMAP and \textit{Planck} data analysis for parity-even bispectra. From the figure we can observe that, in all three models, the polynomial decompositions achieve more than 90\% correlations already for $n_{\rm max} \gtrsim 70$, while the plane-wave decomposed bispectra are less correlated with the theoretical templetes than the polynomial ones. The convergence speed seems to depend strongly on the bispectrum shape. The polynomial basis can reconstruct the pseudoscalar bispectrum more rapidly, achieving 98\% correlation for $n_{\rm max} \simeq 40$. In another equilateral-type case, namely the Weyl model, the convergence is a bit weaker and the correlation reaches around 0.98 for $n_{\rm max} \simeq 100$. For the squeezed-type helical PMF model the convergence is even slower, but we can still get the correlation to exceed 0.98 for $n_{\rm max} \gtrsim 200$. A similar tendency can be seen in the plane-wave decompositions, but the values of correlations drop overall. So we can conclude that for these templates a polynomial decomposition is more efficient than a plane-wave one.

%##################################### 
\subsection{Non-Gaussian map simulation}

We now discuss the issue of simulation of NG maps including bispectra from each of our three theoretical models. The NG part of the map multipoles can be written in separable form for $(x,y,z) = (1,1,-2)$, in the following way
\begin{eqnarray}
a_{\ell m}^{\rm NG} 
\equiv 
\begin{cases}
a_{\ell m}^{{\rm NG} oo} + a_{\ell m}^{{\rm NG} ee} & (\ell = {\rm odd}) \\ 
a_{\ell m}^{{\rm NG} oe} + a_{\ell m}^{{\rm NG} eo} & (\ell = {\rm even})
\end{cases} ~,  \label{eq:almNG_11-2}
\end{eqnarray}
where
\begin{eqnarray}
a_{\ell m}^{{\rm NG} ab}
&=& 
\frac{i \sqrt{C_{\ell}}}{54 v_{\ell}}
\sum_{n \leftrightarrow ijk} \alpha_{n}^Q \int d^2 \hat{\bf n} \nonumber \\ 
&&\times
\left[ 
q_{i}(\ell) {}_{-1} Y_{\ell m}
\left( {}_{-1}M_{j}^{{\rm G} (a)}~ {}_{2}M_{k}^{{\rm G} (b)}
+ {}_{2}M_{j}^{{\rm G} (a)}~ {}_{-1}M_{k}^{{\rm G} (b)} \right) 
\right. \nonumber \\ 
&&\left.\quad + 
q_{i}(\ell) {}_{2} Y_{\ell m} ~ {}_{-1}M_{j}^{{\rm G} (a)} ~  {}_{-1}M_k^{{\rm G} (b)} 
\right. \nonumber \\ 
&&\left.\quad +
q_{j}(\ell) {}_{-1} Y_{\ell m}  
\left(  {}_{-1}M_{k}^{{\rm G} (a)}~ {}_{2}M_{i}^{{\rm G} (b)} 
+ {}_{2}M_{k}^{{\rm G} (a)}~ {}_{-1}M_{i}^{{\rm G} (b)} \right) 
\right. \nonumber \\ 
&&\left.\quad +
q_{j}(\ell) {}_{2} Y_{\ell m} ~ {}_{-1}M_k^{{\rm G} (a)} ~ {}_{-1}M_i^{{\rm G} (b)} 
\right. \nonumber \\ 
&&\left.\quad
+ q_{k}(\ell) {}_{-1} Y_{\ell m} 
\left( 
{}_{-1}M_{i}^{{\rm G} (a)}~ {}_{2}M_j^{{\rm G} (b)} 
+ {}_{2}M_{i}^{{\rm G} (a)}~ {}_{-1}M_j^{{\rm G} (b)}  \right) 
\right. \nonumber \\ 
&&\left.\quad
+ q_{k}(\ell) {}_{2} Y_{\ell m}~  {}_{-1}M_i^{{\rm G} (a)}~ {}_{-1}M_j^{{\rm G} (b)} 
\right]^* 
 ~,
\end{eqnarray}
with $a,b \in o,e$. 

\begin{figure}
  \begin{center}
    \includegraphics[width =1\textwidth]{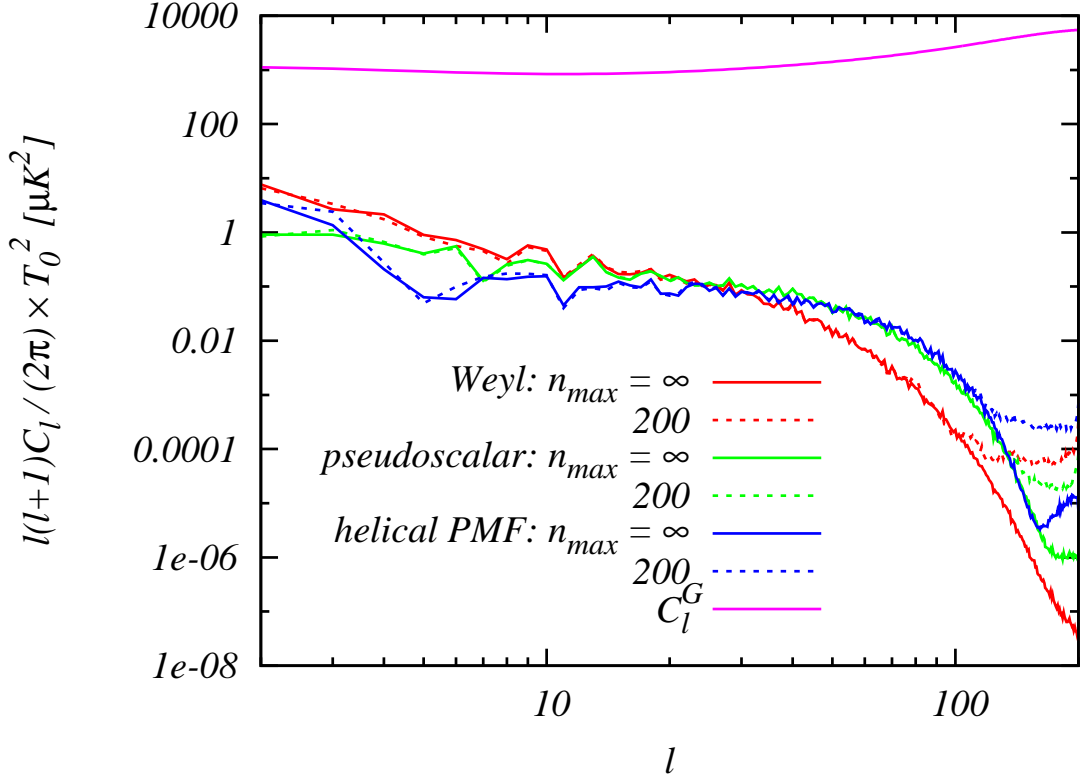}
  \end{center}
  \caption{Power spectra of $a_{\ell m}^{\rm G}$'s in the Weyl (red lines), pseudoscalar (green lines) and helical PMF (blue lines) models, respectively. Here, for comparison of convergence, we plot the exact results, namely $n_{\rm max} = \infty$, estimated in the direct non-separable computations (solid lines), and the results obtained in the $n_{\rm max} = 200$ modal decompositions (dotted lines). $C_\ell^{\rm G}$ denotes the Gaussian part of $C_\ell$.}
  \label{fig:Cl}
\end{figure}

Figure~\ref{fig:Cl} describes the power spectrum of $a_{\ell m}^{\rm NG}$ in each theoretical model. Here we can compare the results from the separable modal decompositions for $n_{\rm max} = 200$ given by eq.~(\ref{eq:almNG_11-2}) with the exact results  from direct non-separable computations of eq.~(\ref{eq:almNG_def}), which should be equivalent to the $n_{\rm max} = \infty$ modal decomposition. As expected, we find that, with pre-computed modal coefficients $\alpha$, the separable modal approach drastically reduces the CPU time from 120 to 0.2 CPU hours. From figure \ref{fig:Cl}, it is also clear that the modal results are in good agreement with the exact non-separable computation up to $\ell \simeq 100$ in all three models. This comes from the fact that, for $n_{\rm max} = 200$, the modal bispectrum reconstructs the exact shape with more than 98\% correlation in each model. In contrast, for $\ell \gtrsim 100$, the modal results tend to deviate from the exact ones. This is due to numerical instabilities in the computation of ${}_{x}M_{i}^{{\rm G} (o/e)}$, and of the angular integrals of their products. We found that this issue can be circumvented by expanding the bispectra and generating maps with an angular resolution much larger than the required $\ell_{\rm max}$ for the analyis, and then smoothing the maps by picking only multipoles up to $\ell_{max}$. Concerning the shapes that are specifically under study, all three spectra decay for $\ell > 100$ due to the end of the tensor-mode ISW enhancement, so this approach is not unfeasible. 

\begin{figure}
\begin{tabular}{c}
    \begin{minipage}{1.0\hsize}
  \begin{center}
    \includegraphics[width = 0.8\textwidth]{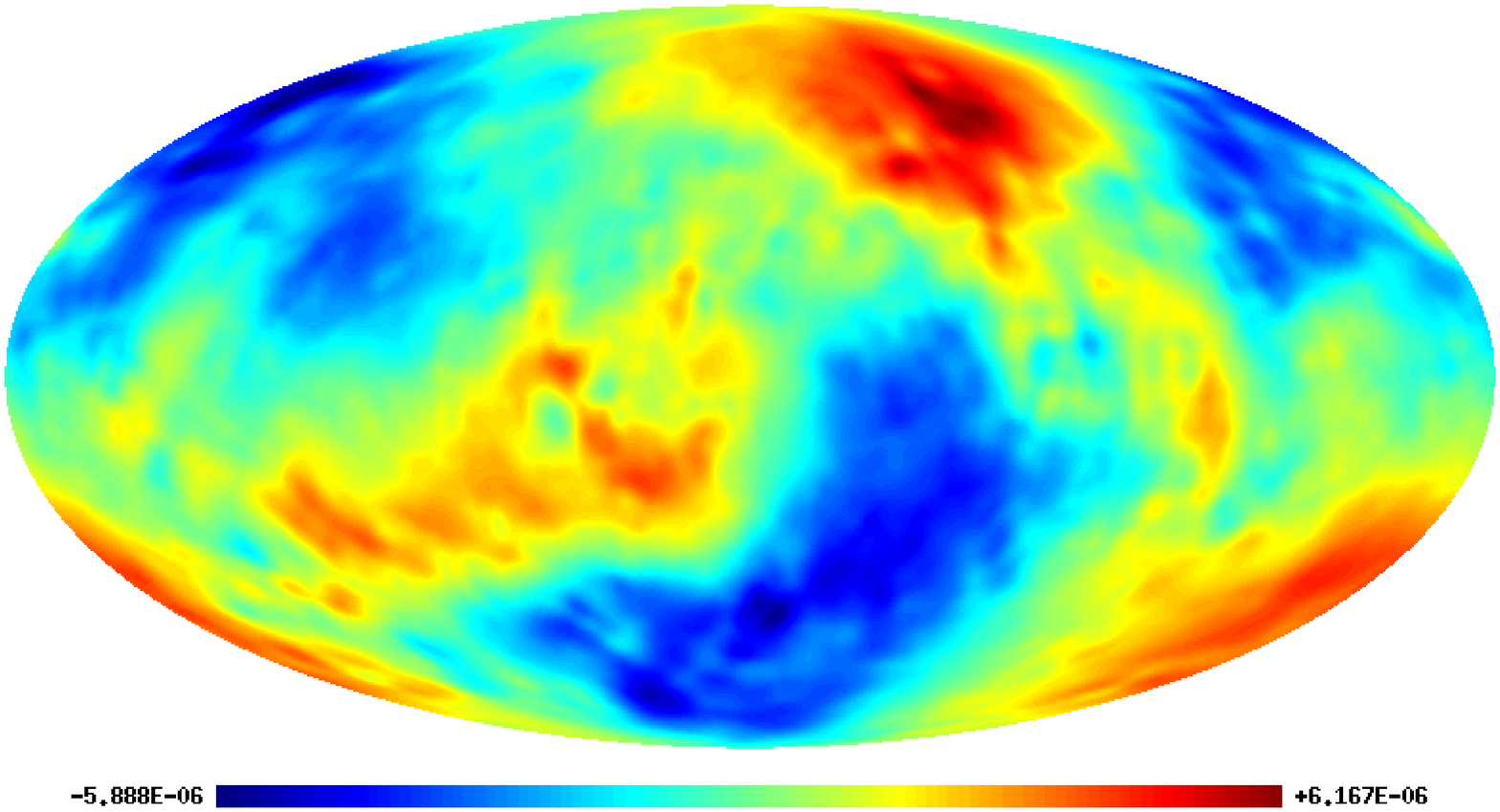}
  \end{center}
\end{minipage}
\end{tabular}
\\
\begin{tabular}{c}
    \begin{minipage}{1.0\hsize}
  \begin{center}
    \includegraphics[width = 0.8\textwidth]{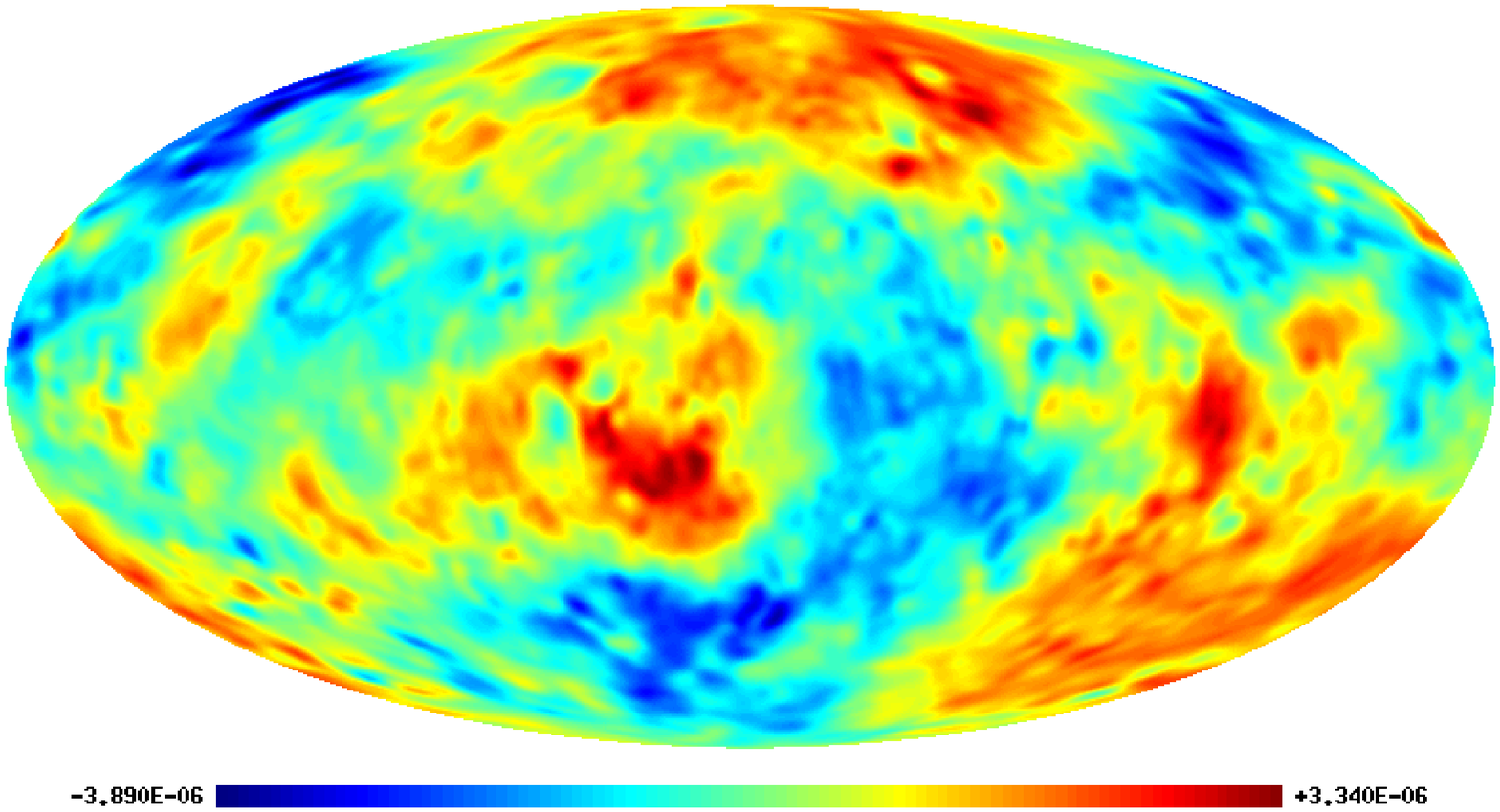}
  \end{center}
\end{minipage}
\end{tabular}
\\
  \begin{tabular}{c}
    \begin{minipage}{1.0\hsize}
  \begin{center}
    \includegraphics[width = 0.8\textwidth]{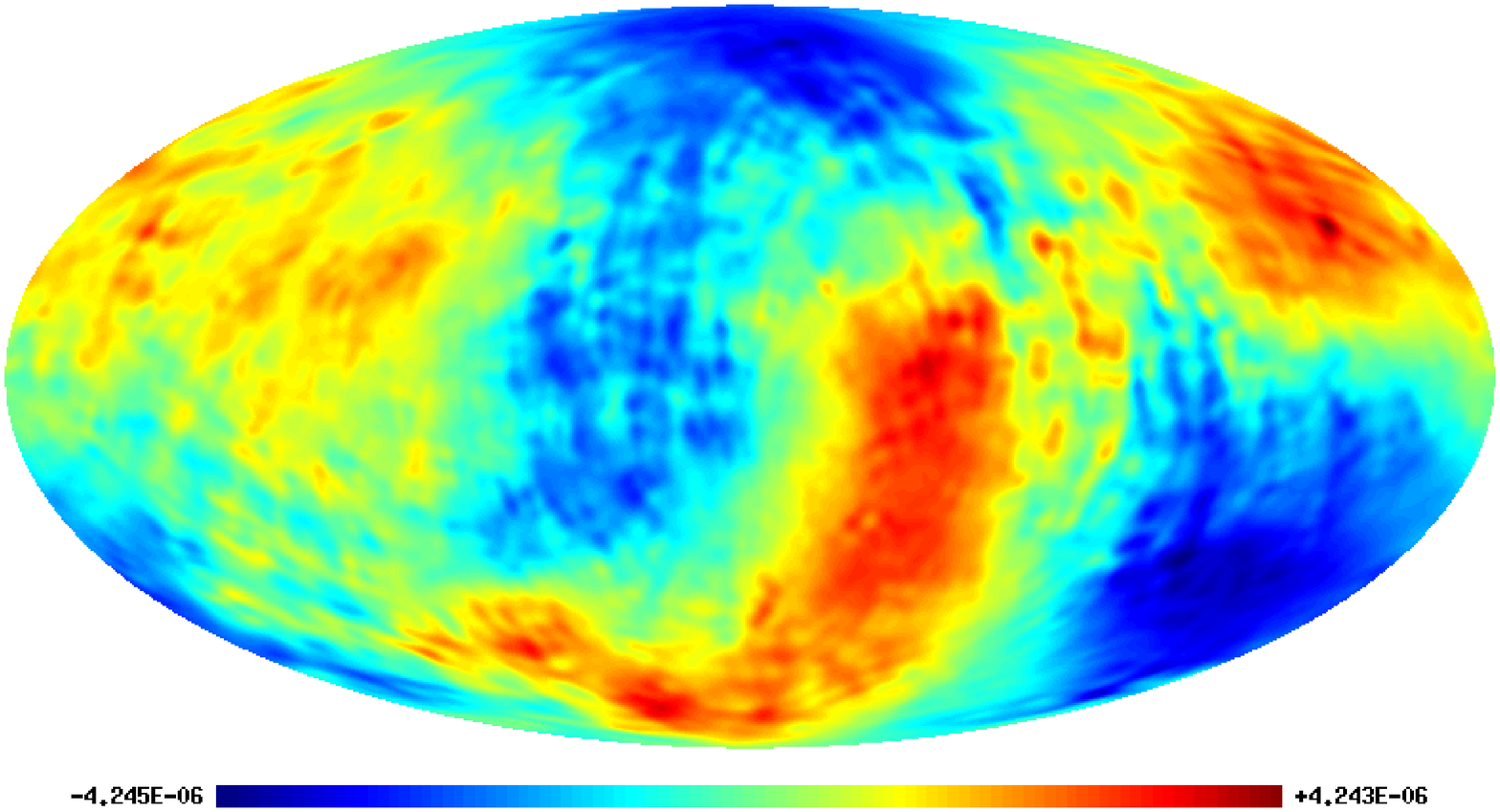}
  \end{center}
\end{minipage}
\end{tabular}
  \caption{Simulated parity-odd NG CMB maps [kelvin] from the Weyl (top panel), pseudoscalar (middle panel) and helical PMF (bottom panel) models, respectively. The maps are generated from $a_{\ell m}^{\rm NG}$'s up to $\ell = 100$, whose power spectra are depicted in figure~\ref{fig:Cl}.}
  \label{fig:map}
\end{figure}

Figure~\ref{fig:map} shows the NG part of a simulated map for each of the three models we are testing; the three maps have been obtained starting from the same Gaussian seed, and we have considered multipoles up to $\ell = 100$, in order to avoid the numerical instabilities mentioned above.

%%%%%%%%%%%%%%%%%%%%%%%%%%%%%%%%%%%%%%%
\section{Conclusion}

Despite the fact that there are several theoretical primordial scenarios predicting the existence of parity-odd bispectra, no observational constraint on this type of NG has been placed so far. Generally, parity-odd bispectra are written in non-separable form, and this has made data analysis unpractical, due to large CPU-time requirements. This paper has developed a new framework for parity-odd CMB bispectrum estimation by extending the separable modal decomposition methodology, already developed and used for parity-even analyses, to parity-odd domains. The analytical extension to the case of interest has been obtained by defining a new reduced bispectrum and a new inner product weight function, in such a way as to account for spin dependence, and to change selection rules in order to include $\ell_1 + \ell_2 + \ell_3 = {\rm odd}$ configurations. In this way, we  
can achieve separability of parity-odd NG estimators and get fast NG maps algorithm, in strict analogy to the parity-even modal expansion procedure. 

Our parity-odd modal decomposition has been numerically implemented and tested by expanding temperature bispectra predicted by several parity-odd Early Universe models. We have checked that the numerical algorithm is stable and achieves convergence using a reasonable number of templates in a few CPU-hours. The exact convergence efficiency depends on the bispectrum shape and the type of modal eigenfunctions. Using decomposed separable bispectra, we have also produced NG simulations and checked the consistency with the exact results from a slow brute-force approach. As expected, we get massive computational gains when working with separable modal bispectra. 

The algorithm for bispectrum estimation developed in this paper is applicable to all types of parity-odd bispectra (i.e., bispectra enforcing the condition $\ell_1 + \ell_2 + \ell_3 = {\rm odd}$). Our numerical approach so far has included only temperature bispectra. 
Future interesting applications will include actual estimation of parity-odd NG from CMB data, and the extension of our method to polarized bispectra, which are generally predicted in parity-odd scenarios, with taking care of bias due to the experimental systematics or the imperfect sky coverage.

%%%%%%%%%%%%%%%%%%%%%%%%%%%%%%%%%%%%%%% 
\acknowledgments
We are very grateful to Paul Shellard for many useful discussions. MS is supported in part by a Grant-in-Aid for JSPS Research under Grant No.~25-573. This work is supported in part by the ASI/INAF Agreement I/072/09/0 for the Planck LFI Activity of Phase E2.

%\appendix

%%%%%%%%%%%%%%%%%%%%%%%%%%%%%%%%%%%%%%%

%%%%%%%%%%%%%%%%%%%%%%%%%%%%%%%%%%%%%%%%
\bibliography{paper}
\end{document}